\documentclass[final,3p,times,twocolumn]{elsarticle}
\usepackage{epsfig}
\usepackage{amsmath}
\usepackage{color}
\usepackage{latexsym}
\usepackage{amssymb}
\usepackage{dsfont}
\usepackage{multirow}

\newcommand{\bea}{\begin{eqnarray}}
\newcommand{\eea}{\end{eqnarray}}
\newcommand{\be}{\begin{equation}}
\newcommand{\ee}{\end{equation}}

\newcommand{\ud}{\mathrm{d}}

\newcommand{\barpsi}{\overline{\psi}}

\newlength\savedwidth

\begin{document}

\title{Wilson lines and orbital angular momentum}

\author{C\'edric Lorc\'e}
\address{IPNO, Universit\'e Paris-Sud, CNRS/IN2P3, 91406 Orsay, France\\
        and LPT, Universit\'e Paris-Sud, CNRS, 91406 Orsay, France}

\begin{abstract}
We present an explicit realization of the Chen \emph{et al.} approach to the proton spin decomposition in terms of Wilson lines, generalizing the light-front gauge-invariant extensions discussed recently by Hatta. Particular attention is drawn to the residual gauge freedom by further separating the pure-gauge term into contour and residual terms. We show that the kinetic orbital angular momentum operator can be expressed in terms of the Wigner operator only when the momentum variable is integrated over. Finally, we confirm from twist-2 arguments that the advanced, retarded and antisymmetric light-front canonical orbital angular momenta are the same. 
\end{abstract}

\maketitle

\section{Introduction}

In the last few years, the debates about which decomposition of the proton spin is physically acceptable were revived by the possibility of rendering the Jaffe-Manohar or canonical decomposition gauge invariant. Moreover, it has been shown that one can access the canonical orbital angular momentum provided that one is able to extract experimentally either the Wigner distributions or particular twist-3 distributions. For a recent review of the discussions, see Ref.~\cite{Lorce:2012rr}. 

This work addresses the issues of explicit definition, uniqueness and measurability of the canonical decomposition from a geometrical perspective. In section \ref{sec2}, we first recall the suggestion made by Chen \emph{et al.} to separate the gauge field into pure-gauge and physical terms. Although gauge invariant, this approach is not unique owing to the Stueckelberg symmetry which reflects the freedom in defining what is exactly meant by pure-gauge and physical contributions.  In section \ref{sec3}, we present a generic realization of the Chen \emph{et al.} approach based on the idea of parallel transport from a reference point. We show in particular that a change of reference point and/or path amounts to a Stueckelberg transformation. We discuss with special care in section \ref{sec4} the partial gauge fixing and the residual gauge freedom which motivate a further decomposition of the pure-gauge term into contour and residual terms. In section \ref{sec5}, we show that the Wigner operator is naturally related to the canonical momentum operator. It can also be related to the kinetic momentum operator provided that one integrates over the momentum variable. Choosing Wilson lines running along the light-front direction, we recover in section \ref{sec6} the light-front gauge-invariant extensions of the canonical angular momentum discussed in the literature, albeit with a more transparent treatment of the residual gauge freedom. Then we show from twist-2 arguments that the advanced, retarded and antisymmetric light-front canonical orbital angular momenta are the same, confirming a previous approach based on the twist-3 level. Finally, we gather our conclusions in section \ref{sec7}.

\section{Chen \emph{et al.} decomposition}\label{sec2}

In order to unambiguously define what is meant by gluon spin and orbital angular momentum, Chen \emph{et al.} proposed to split the gauge field into a pure-gauge term and a physical term \cite{Chen:2008ag,Chen:2011zzh,Wakamatsu:2010qj,Wakamatsu:2010cb}
\begin{equation}\label{Adecomp}
A_\mu(x)=A^\text{pure}_\mu(x)+A^\text{phys}_\mu(x)
\end{equation}
satisfying specific gauge transformation laws
\begin{align}
A_\mu^\text{pure}(x)&\mapsto\tilde A_\mu^\text{pure}(x)=\nonumber\\
&\hspace{1cm}U(x)\left[A_\mu^\text{pure}(x)+\frac{i}{g}\,\partial_\mu\right]U^{-1}(x),\label{Apureg}\\
A_\mu^\text{phys}(x)&\mapsto\tilde A_\mu^\text{phys}(x)=U(x)A_\mu^\text{phys}(x)U^{-1}(x).\label{Aphysg}
\end{align}
Since $A^\text{pure}_\mu(x)$ is a pure gauge, it can be written as
\begin{equation}
A_\mu^\text{pure}(x)=\frac{i}{g}\,U_\text{pure}(x)\partial_\mu U^{-1}_\text{pure}(x),
\end{equation}
where $U_\text{pure}(x)$ is some unitary gauge matrix  with the gauge transformation law
\begin{equation}\label{Ugauge}
U_\text{pure}(x)\mapsto\tilde U_\text{pure}(x)=U(x)U_\text{pure}(x).
\end{equation}
Clearly, in the gauge $U(x)=U^{-1}_\text{pure}(x)$ the pure-gauge term vanishes.

By construction, the decomposition \eqref{Adecomp} is gauge invariant. However, it is not unique since we still have some freedom in defining exactly what we mean by `pure-gauge' and `physical'. The reason is that the pure-gauge and physical terms remain respectively pure-gauge and physical under the following transformation which leaves $A_\mu(x)$ invariant
\begin{align}
A^\text{pure}_\mu(x)&\mapsto A^{\text{pure},g}_\mu(x)=\nonumber\\
&\hspace{-0.75cm}A^\text{pure}_\mu(x)+\frac{i}{g}\,U_\text{pure}(x)U_0^{-1}(x)\left[\partial_\mu U_0(x)\right]U^{-1}_\text{pure}(x),\label{stueck1}\\
A^\text{phys}_\mu(x)&\mapsto A^{\text{phys},g}_\mu(x)=\nonumber\\
&\hspace{-0.75cm}A^\text{phys}_\mu(x)-\frac{i}{g}\,U_\text{pure}(x)U_0^{-1}(x)\left[\partial_\mu U_0(x)\right]U^{-1}_\text{pure}(x),\label{stueck2}
\end{align}
where $U_0(x)$ is a gauge-invariant unitary matrix. At the level of $U_\text{pure}(x)$, this transformation reads
\begin{equation}\label{Ustueck}
U_\text{pure}(x)\mapsto U^g_\text{pure}(x)=U_\text{pure}(x)U^{-1}_0(x).
\end{equation}
While the ordinary gauge transformation acts on the left of $U_\text{pure}(x)$ as in Eq. \eqref{Ugauge}, this new transformation acts on the right. It is therefore important to distinguish them. Because of the similarity between the Chen \emph{et al.} approach and the Stueckelberg mechanism \cite{Lorce:2012rr}, we refer to this transformation as the Stueckelberg (gauge) transformation.

\section{Explicit construction using Wilson lines}\label{sec3}

In this section, we present an explicit construction of the Chen \emph{et al.} decomposition based on the idea of parallel transport. In general relativity the fiber, \emph{i.e.} the vector space attached to each point $x$, is the tangent space which inherits automatically a natural basis. Indeed, once a system of coordinates $x^\mu$ is chosen in a region of space-time, the natural basis in the tangent space is given by the derivatives with respect to the coordinates $\partial_\mu$. So the same index $\mu$ can refer either to a coordinate in space-time or to a component of an object living in the tangent space like \emph{e.g.} a directional derivative. Fixing some point and some tangent vector at that point, the parallel transport equation can then be used to define a unique curve in space-time called the geodesic. 

In gauge theories, the fiber is an internal space. Fixing a basis in this internal space, one can work with the components in internal space. For example, the quark field $\psi^a(x)$ at space-time point $x$ has one internal index $a$. A gauge transformation is nothing but a change of basis in internal space, which can be different for each space-time point $\psi^a(x)\mapsto\tilde\psi^a(x)=U^a_{\phantom{a}b}(x)\psi^b(x)$. The gauge symmetry simply means that the physics does not depend on the arbitrary choice of basis in internal space. Consequently, contrary to general relativity, there is \emph{a priori} no natural (internal) basis in gauge theories.

Our aim is to \emph{define}, for each space-time point, a natural basis in internal space. The idea is to fix a natural basis at some reference point, and then use the parallel transport equation to single out the natural basis associated to any other point. Since the parallel transport depends on the path followed to connect two points, we have also to define (arbitrarily) the ``shape'' of the contour. Such contours are called in the following \emph{standard} contours. The most important constraints in the choice of the standard contours are the absence of self-intersection and the existence of a standard contour connecting the reference point to any other point \cite{Ivanov:1985np,Shevchenko:1998uw}.

Consider some reference point $x_r$. At that point, we can (arbitrarily) define in internal space both a natural basis and the actual basis for the calculation. The elements of the matrix $U_\text{pure}(x_r)$ simply give the components of the actual basis vectors in the natural basis. Then, we can parallel transport $U_\text{pure}(x_r)$ to any other point $x$ along a standard contour $\mathcal C$ parametrized by the path $s(\lambda)$, and therefore define a unique $U_\text{pure}(x)$. As usual, the parallel transport equation expresses the fact that the covariant derivative along the path has to vanish
\begin{equation}
\frac{\partial s^\mu}{\partial \lambda}\,D_\mu U_\text{pure}(s(\lambda))=0,
\end{equation}
where $D_\mu=\partial_\mu-igA_\mu(s(\lambda))$ is the covariant derivative. The solution to this equation involves the well-known Wilson line
\begin{align}
U_\text{pure}(x)&=\mathcal W_{\mathcal C}(x,x_r)\,U_\text{pure}(x_r),\\
\mathcal W_{\mathcal C}(x,x_r)&=\mathcal P\left[e^{ig\int_{x_r}^xA_\mu(s)\,\ud s^\mu}\right]\equiv\mathds 1+ig\int_{x_r}^xA_\mu(s)\,\ud s^\mu\nonumber\\
&\hspace{-1cm}+(ig)^2\int_{x_r}^x\int_{x_r}^{s_1}A_\mu(s_1)A_\nu(s_2)\,\ud s^\mu_1\,\ud s^\nu_2+\cdots.
\end{align}
Using now the formula for the derivative of the Wilson line
\begin{align}
\frac{\partial}{\partial z^\mu}\mathcal W_{\mathcal C}(x,y)&=ig\,\mathcal W_{\mathcal C}(x,s)\,A_\alpha(s)\,\frac{\partial s^\alpha}{\partial z^\mu}\,\mathcal W_{\mathcal C}(s,y)\Big|_{s=y}^{s=x}\nonumber\\
&\hspace{-1.5cm}+ig\int_y ^x\mathcal W_{\mathcal C}(x,s)\,F_{\alpha\beta}(s)\,\mathcal W_{\mathcal C}(s,y)\,\frac{\partial s^\alpha}{\partial z^\mu}\,\ud s^\beta,\label{derivativeW}
\end{align}
where $F_{\alpha\beta}=\partial_\alpha A_\beta-\partial_\beta A_\alpha-ig\left[A_\alpha,A_\beta\right]$ is the field-strength tensor, and the fact that the inverse of the Wilson line is simply obtained by an interchange of the end points $\mathcal W^{-1}_{\mathcal C}(x,y)=\mathcal W_{\mathcal C}(y,x)$, we arrive at the following explicit expressions for the pure-gauge and physical parts of the gauge field
\begin{align}
A^\text{pure}_\mu(x)&=\nonumber\\
&\hspace{-1cm}\mathcal W_{\mathcal C}(x,x_r)\left[\frac{\partial x^\alpha_r}{\partial x^\mu}\,A^\text{pure}_\alpha(x_r)+\frac{i}{g}\,\frac{\partial}{\partial x^\mu}\right]\mathcal W_{\mathcal C}(x_r,x),\label{Apurex}\\
A^\text{phys}_\mu(x)&=\mathcal W_{\mathcal C}(x,x_r)\,\frac{\partial x^\alpha_r}{\partial x^\mu}\,A^\text {phys}_\alpha(x_r)\,\mathcal W_{\mathcal C}(x_r,x)\nonumber\\
&\hspace{-1cm}-\int_{x_r}^x \mathcal W_{\mathcal C}(x,s)\,F_{\alpha\beta}(s)\,\mathcal W_{\mathcal C}(s,x)\,\frac{\partial s^\alpha}{\partial x^\mu}\,\ud s^\beta.\label{Aphysx}
\end{align}
Eqs. \eqref{Apurex} and \eqref{Aphysx} are nothing but the parallel transport of $A^\text{pure}_\mu (x_r)$ and $A^\text{phys}_\mu(x_r)$ to the point $x$. Using the gauge transformation law of the Wilson line
\begin{equation}\label{Wilsong}
\mathcal W_{\mathcal C}(x,y)\mapsto\tilde{\mathcal W}_{\mathcal C}(x,y)=U(x)\,\mathcal W_{\mathcal C}(x,y)\,U^{-1}(y),
\end{equation}
it is straightforward to check that the pure-gauge and physical terms in Eqs. \eqref{Apurex} and \eqref{Aphysx} transform according to Eqs. \eqref{Apureg} and \eqref{Aphysg}, respectively. 

Since all the points belonging to a standard contour have the same reference point $\frac{\partial x^\alpha_r}{\partial\lambda}=0$, the physical part of the gauge potential is orthogonal to the contour
\begin{equation}
 \frac{\partial s^\mu}{\partial\lambda}\,A^\text{phys}_\mu(s(\lambda))=0.
\end{equation}
In other words, the component of the gauge field tangent to the path is considered to be a pure gauge. This clearly shows how the choice of a standard contour affects the separation of the gauge field into pure-gauge and physical terms. By simply changing the contour and possibly the reference point, we change what we mean by pure-gauge and physical terms, \emph{i.e.} such a change amounts to performing a Stueckelberg transformation. It is easy to relate different choices of contour and reference point. Indeed, denoting the fields obtained with a different contour $\mathcal C'$ and possibly different reference point $x'_r$ with a superscript $g$, we can write
\begin{equation}\label{stueck3}
U^g_\text{pure}(x)=U_\text{pure}(x)\,U^{-1}_0(x),
\end{equation}
where $U_0(x)=U^{g,-1}_\text{pure}(x'_r)\,\mathcal W_{\mathcal C'}(x'_r,x)\,\mathcal W_{\mathcal C}(x,x_r)\,U_\text{pure}(x_r)$ is obviously unitary and gauge invariant. Eq. \eqref{stueck3} is then nothing but the Stueckelberg transformation \eqref{Ustueck}.

\section{Contour, residual and natural gauges}\label{sec4}

The contour gauge corresponds to the gauge where the Wilson line reduces to the identity matrix for any $x$. This amounts to taking $U(x)=\mathcal W_{\mathcal C}(x_r,x)$. Contour gauges are particularly interesting because they were shown to be ghost-free \cite{Ivanov:1985np}. Denoting the fields in the contour gauge with a superscript $\mathcal C$, we have
\begin{align}
A^{\text{pure},\mathcal C}_\mu(x)&=\frac{\partial x^\alpha_r}{\partial x^\mu}\,A^\text{pure}_\alpha(x_r),\label{ApureC}\\
A^{\text{phys},\mathcal C}_\mu(x)&=\frac{\partial x^\alpha_r}{\partial x^\mu}\,A^\text {phys}_\alpha(x_r)-\int_{x_r}^x F^{\mathcal C}_{\alpha\beta}(s)\,\frac{\partial s^\alpha}{\partial x^\mu}\,\ud s^\beta.\label{AphysC}
\end{align}
Note in particular that $A^{\mathcal C}_\mu(x_r)=A_\mu(x_r)$. The reason for this is simply that we used the point $x_r$ as a reference, and so the original field and the field in the contour gauge should coincide at that point.

In general, the reference point $x_r$ depends on $x$. Gauge transformations depending only implicitly on $x$ through $x_r$ will then leave the Wilson line in the contour gauge $\mathcal W^{\mathcal C}_{\mathcal C}(x,x_r)=\mathds 1$ invariant.
This remaining arbitrariness is nothing but the residual gauge symmetry
\begin{align}
A^{\text{pure},\mathcal C}_\mu(x)&\mapsto \tilde A^{\text{pure},\mathcal C}_\mu(x)=\nonumber\\
&\quad U(x_r)\left[A^{\text{pure},\mathcal C}_\mu(x)+\frac{i}{g}\,\frac{\partial}{\partial x^\mu}\right]U^{-1}(x_r),\label{Apureres}\\
A^{\text{phys},\mathcal C}_\mu(x)&\mapsto \tilde A^{\text{phys},\mathcal C}_\mu(x)=\nonumber\\
&\quad U(x_r)\,A^{\text{phys},\mathcal C}_\mu(x)\,U^{-1}(x_r).\label{Aphysres}
\end{align}
It seems therefore natural to split further the pure-gauge term as follows
\begin{equation}
A^\text{pure}_\mu(x)=A^\text{con}_\mu(x)+A^\text{res}_\mu(x),
\end{equation}
where the contour and residual terms are defined as
\begin{align}
A^\text{con}_\mu(x)&=\frac{i}{g}\,\mathcal W_{\mathcal C}(x,x_r)\,\frac{\partial}{\partial x^\mu}\mathcal W_{\mathcal C}(x_r,x),\\
A^\text{res}_\mu(x)&=\mathcal W_{\mathcal C}(x,x_r)\,\frac{\partial x^\alpha_r}{\partial x^\mu}\,A^\text{pure}_\alpha(x_r)\,\mathcal W_{\mathcal C}(x_r,x).
\end{align}
Their gauge transformation laws can easily be derived from Eq. \eqref{Wilsong}
\begin{align}
A^\text{con}_\mu(x)&\mapsto \tilde A^\text{con}_\mu(x)
=U(x)\left[A^\text{con}_\mu(x)+\frac{i}{g}\,\partial_\mu\right]U^{-1}(x)\nonumber\\
&\hspace{-0.75cm}+\frac{i}{g}\,U_{\mathcal C}(x,x_r)\,U^{-1}(x_r)\left[\frac{\partial}{\partial x^\mu}U(x_r)\right]U^{-1}_{\mathcal C}(x,x_r),\\
A^\text{res}_\mu(x)&\mapsto \tilde A^\text{res}_\mu(x)
=U(x)\,A^\text{res}_\mu(x)U^{-1}(x)\nonumber\\
&\hspace{-0.75cm}-\frac{i}{g}\,U_{\mathcal C}(x,x_r)\,U^{-1}(x_r)\left[\frac{\partial}{\partial x^\mu}U(x_r)\right]U^{-1}_{\mathcal C}(x,x_r),
\end{align}
where $U_{\mathcal C}(x,x_r)\equiv U(x)\,\mathcal W_{\mathcal C}(x,x_r)$. Clearly, in the contour gauge, the contour term vanishes $A^{\text{con},\mathcal C}_\mu(x)=0$ and is invariant under residual gauge transformations, so that the pure-gauge term simply reduces to the residual term $A^{\text{pure},\mathcal C}_\mu(x)=A^{\text{res},\mathcal C}_\mu(x)$. Note also that in order to define the physical part of the gauge potential at a point $x$, we need to know what is the physical part at the reference point $x_r$. It is therefore more natural to consider the sum of physical and residual parts, as its definition involves solely the full gauge field $A_\mu(x)$
\begin{align}
A^\text{res}_\mu(x)+A^\text{phys}_\mu(x)&=\mathcal W_{\mathcal C}(x,x_r)\,\frac{\partial x^\alpha_r}{\partial x^\mu}\,A_\mu(x_r)\,\mathcal W_{\mathcal C}(x_r,x)\nonumber\\
&\hspace{-1.5cm}-\int_{x_r}^x \mathcal W_{\mathcal C}(x,s)\,F_{\alpha\beta}(s)\,\mathcal W_{\mathcal C}(s,x)\,\frac{\partial s^\alpha}{\partial x^\mu}\,\ud s^\beta.
\end{align}
In the literature, see \emph{e.g.} Ref. \cite{Hatta:2011ku}, this sum is sometimes also referred to as $A^\text{phys}_\mu(x)$, introducing some confusion in what is meant by physical. Strictly speaking, even though the sum $A^\text{res}_\mu(x)+A^\text{phys}_\mu(x)$ is orthogonal to the contour, it does not transform covariantly under gauge transformations \eqref{Aphysg}, and should not be considered as a genuine physical part.

As discussed in Ref.~\cite{Lorce:2012rr}, the natural gauge is the gauge where the pure-gauge term vanishes, and corresponds to taking $U(x)=U^{-1}_\text{pure}(x)$. Equivalently, we can start with the fields in the contour gauge and perform a residual gauge transformation with $U(x_r)=U^{-1}_\text{pure}(x_r)$. Denoting the fields in the natural gauge with a hat, we obtain
\begin{align}
\hat A^\text{pure}_\mu(x)
&=\hat A^\text{con}_\mu(x)=\hat A^\text{res}_\mu(x)=0,\\
\hat A^\text{phys}_\mu(x)&=\frac{\partial x^\alpha_r}{\partial x^\mu}\,\hat A_\alpha(x_r)-\int_{x_r}^x \hat F_{\alpha\beta}(s)\,\frac{\partial s^\alpha}{\partial x^\mu}\,\ud s^\beta.
\end{align}
In practice, one would often assume that all the standard contours start at the same fixed point $x_0$. Choosing this fixed point as the reference point $x_r=x_0$, one avoids dealing with the residual gauge symmetry
\begin{align}
U_\text{pure}(x)&=\mathcal W_{\mathcal C}(x,x_0)U_\text{pure}(x_0),\\
A^\text{pure}_\mu(x)&=\frac{i}{g}\,\mathcal W_{\mathcal C}(x,x_0)\,\frac{\partial}{\partial x^\mu}\mathcal W_{\mathcal C}(x_0,x),\label{Apurefixed}\\
A^\text{phys}_\mu(x)&=-\int_{x_0}^x \mathcal W_{\mathcal C}(x,s)\,F_{\alpha\beta}(s)\,\mathcal W_{\mathcal C}(s,x)\,\frac{\partial s^\alpha}{\partial x^\mu}\,\ud s^\beta,\label{comment}
\end{align}
and so the contour gauge simply coincides with the natural gauge up to a global (\emph{i.e.} $x$-independent) gauge transformation. Note also from Eq. \eqref{comment} that the physical part of the gauge field vanishes when $x=x_0$, which implies the pointwise equality $A_\mu(x_0)=A^\text{pure}_\mu(x_0)$. However, one cannot conclude that $F_{\mu\nu}(x_0)=0$ since the field-strength tensor involves also the derivatives of the gauge field. 

Consider now that the reference point is some intermediate point on the path from $x_0$ to $x$. We can decompose the parallel transport from $x_0$ to $x$ as a parallel transport from $x_0$ to $x_r$, followed by a parallel transport from $x_r$ to $x$
\begin{equation}
\mathcal W_{\mathcal C}(x,x_0)=\mathcal W_{\mathcal C}(x,x_r)\,\mathcal W_{\mathcal C}(x_r,x_0).
\end{equation}
The parallel transport from $x_0$ to $x_r$ defines what is the pure-gauge term at $x_r$. Choosing $x_r$ as the reference point leads to the identification of the pure-gauge term at $x_r$ with the residual term $A^\text{pure}_\mu(x_r)=A^\text{res}_\mu(x_r)$. The parallel transport from $x_r$ to $x$ then defines what is the contour term. In short, the parallel transport from $x_0$ defines what is the pure-gauge term, while the parallel transport from $x_r$ determines the separation of the pure-gauge term into contour and residual terms.

\section{Orbital angular momentum in phase space}\label{sec5}

Since the orbital angular momentum (OAM) corresponds to a correlation between position and momentum, it is most easily discussed from a phase-space perspective. In this section, we restrict ourselves to the quark sector, but the discussion can easily be transposed to the gluon sector. We first remind the two main kinds of OAM and the definition of the gauge-invariant Wigner or phase-space operator. We then show how this Wigner operator is related to the OAM. 

\subsection{Kinetic and canonical orbital angular momentum}

There exist essentially two kinds of gauge-invariant orbital angular momentum. One is the \emph{kinetic} OAM \cite{Ji:1996ek}
\begin{equation}
\mathcal M^{\mu\nu\rho}_{q,\text{OAM}}(x)=\frac{i}{2}\,\barpsi(x) \gamma^\mu x^{[\nu}\!\!\stackrel{\leftrightarrow}{D}\!\!\!\!\!\phantom{\partial}^{\rho]}(x)\psi(x)
\end{equation}
and the other one is the \emph{canonical} OAM \cite{Chen:2008ag,Chen:2011zzh}
\begin{equation}\label{canonicalOAM}
\mathsf M^{\mu\nu\rho}_{q,\text{OAM}}(x)=\frac{i}{2}\,\barpsi(x) \gamma^\mu x^{[\nu}\!\!\stackrel{\leftrightarrow}{D}\!\!\!\!\!\phantom{\partial}^{\rho]}_\text{pure}(x)\psi(x),
\end{equation}
where the covariant derivatives at the point $x$ are defined as $D^\mu(x)=\partial^\mu-igA^\mu(x)$ and $D^\mu_\text{pure}(x)=\partial^\mu-igA^\mu_\text{pure}(x)$.  We used for convenience the notations $a^{[\mu}b^{\nu]}=a^\mu b^\nu-a^\nu b^\mu$ and $\stackrel{\leftrightarrow}{\partial}\,=\,\stackrel{\rightarrow}{\partial}-\stackrel{\leftarrow}{\partial}$. These two OAMs differ by a so-called potential term \cite{Wakamatsu:2010qj,Wakamatsu:2010cb}
\begin{equation}
\mathsf M^{\mu\nu\rho}_\text{pot}(x)=-g\,\barpsi(x) \gamma^\mu x^{[\nu}A^{\rho]}_\text{phys}(x)\psi(x),
\end{equation}
which is usually non-vanishing. In the natural gauge, the canonical OAM reduces to the same expression as in the definition of the Jaffe-Manohar OAM \cite{Jaffe:1989jz} 
\begin{equation}
\hat{\mathsf M}^{\mu\nu\rho}_{q,\text{spin}}(x)=\frac{i}{2}\,\hat\barpsi(x) \gamma^\mu x^{[\nu}\!\!\stackrel{\leftrightarrow}{\partial}\!\!\!\!\!\phantom{\partial}^{\rho]}\hat\psi(x),
\end{equation}
and can then be thought of as a gauge-invariant extension (GIE) of the latter \cite{Ji:2012sj,Ji:2012ba}.

Contrarily to the kinetic OAM, the canonical OAM is not Stueckelberg invariant\footnote{In the gluon sector, only the total (\emph{i.e.} spin+OAM) kinetic angular momentum is Stueckelberg invariant.}, \emph{i.e.} it depends on how one explicitly separates the gauge field into pure-gauge and physical terms. There is consequently an infinite number of possible different definitions of canonical OAM, all sharing the same formal structure \eqref{canonicalOAM}. The reduction to the Jaffe-Manohar OAM occurs in different gauges, implying that the different canonical OAMs are not equivalent.

\subsection{Wigner operator}

The gauge-invariant quark Wigner operator is defined as \cite{Ji:2003ak,Belitsky:2003nz}
\begin{align}
\widehat W^{[\gamma^\mu]q}(x,k)&\equiv\int\frac{\ud^4z}{(2\pi)^4}\,e^{ik\cdot z}\nonumber\\
&\hspace{-0.5cm}\barpsi(x-\tfrac{z}{2}) \gamma^\mu\,\mathcal W_{\mathcal C}(x-\tfrac{z}{2},x+\tfrac{z}{2})\,\psi(x+\tfrac{z}{2}).
\end{align}
It can be interpreted as a phase-space density operator. For example, the first and second $k$-moments of the Wigner operator respectively give the density and canonical momentum density operators in coordinate space
\begin{align}
\int\ud^4k\,\widehat W^{[\gamma^\mu]q}(x,k)&=\barpsi(x) \gamma^\mu\psi(x),\\
\int\ud^4k\,k^\rho\,\widehat W^{[\gamma^\mu]q}(x,k)&=\frac{i}{2}\,\barpsi(x) \gamma^\mu \!\!\stackrel{\leftrightarrow}{D}\!\!\!\!\!\phantom{\partial}^\rho_\text{pure}(x)\psi(x).\label{secondmoment}
\end{align}
The first moment is trivially obtained, while the second requires some care. We sketch here its derivation. We first assume that the Wilson line appearing in the definition of the Wigner operator is composed of standard paths, and that there exists a fixed reference point $x_0$. Integrating by parts and using the fact that $x$ and $z$ are independent variables, we can write
\begin{align}
&\int\ud^4k\,k^\rho\,\widehat W^{[\gamma^\mu]q}(x,k)=i\int\frac{\ud^4k\,\ud^4z}{(2\pi)^4}\,e^{ik\cdot z}\nonumber\\
&\hspace{0.75cm}\partial^\rho_z\left[\barpsi(x-\tfrac{z}{2})\gamma^\mu\,\mathcal W_{\mathcal C}(x-\tfrac{z}{2},x+\tfrac{z}{2})\,\psi(x+\tfrac{z}{2})\right]\label{intermediatestep}\\
&=\frac{i}{2}\int\frac{\ud^4k\,\ud^4z}{(2\pi)^4}\,e^{ik\cdot z}\,\barpsi_{\mathcal C}(x-\tfrac{z}{2},x_0)\gamma^\mu\!\!\stackrel{\leftrightarrow}{\partial}\!\!\!\!\!\phantom{\partial}^\rho_x\psi_{\mathcal C}(x_0,x+\tfrac{z}{2}),\nonumber
\end{align}
where $\psi_{\mathcal C}(x,y)\equiv\mathcal W_{\mathcal C}(x,y)\,\psi(y)$ and $\barpsi_{\mathcal C}(x,y)\equiv\barpsi(x)\,\mathcal W_{\mathcal C}(x,y)$. From the definition \eqref{Apurefixed} for $A^\mu_\text{pure}(x)$, we obtain
\begin{align}
&\mathcal W_{\mathcal C}(x-\tfrac{z}{2},x_0)\!\!\stackrel{\leftrightarrow}{\partial}\!\!\!\!\!\phantom{\partial}^\rho_x\mathcal W_{\mathcal C}(x_0,x+\tfrac{z}{2})=\nonumber\\
&\hspace{1.5cm}\mathcal W_{\mathcal C}(x-\tfrac{z}{2},x)\!\!\stackrel{\leftrightarrow}{D}\!\!\!\!\!\phantom{\partial}^\rho_\text{pure}(x)\mathcal W_{\mathcal C}(x,x+\tfrac{z}{2}).\label{skewedmomop}
\end{align}
Inserting now this expression in Eq. \eqref{intermediatestep} and performing the integrations, we arrive at the result \eqref{secondmoment}. It is therefore natural to interpret the variable $k^\mu$ appearing in the definition of the Wigner operator as the (average) canonical momentum.

Is it possible to define a Wigner operator where the variable $k$ would be interpreted as the (average) kinetic momentum? The answer is negative. Indeed, consider any generic differential operator $\stackrel{\leftrightarrow}{\mathcal D}\!\!\!\!\!\phantom{\partial}^\rho(x)$ defined at some point $x$. There exist two natural non-local generalizations of such an operator
\begin{align}
\stackrel{\leftrightarrow}{\mathcal D}\!\!\!\!\!\phantom{\partial}^\rho_m(x-\tfrac{z}{2},x+\tfrac{z}{2})&\equiv\nonumber\\
&\hspace{-0.5cm}\mathcal W_{\mathcal C}(x-\tfrac{z}{2},x)\!\stackrel{\leftrightarrow}{\mathcal D}\!\!\!\!\!\phantom{\partial}^\rho(x)\mathcal W_{\mathcal C}(x,x+\tfrac{z}{2}),\\
\stackrel{\leftrightarrow}{\mathcal D}\!\!\!\!\!\phantom{\partial}^\rho_e(x-\tfrac{z}{2},x+\tfrac{z}{2})&\equiv\mathcal W_{\mathcal C}(x-\tfrac{z}{2},x+\tfrac{z}{2})\!\stackrel{\rightarrow}{\mathcal D}\!\!\!\!\!\phantom{\partial}^\rho(x+\tfrac{z}{2})\nonumber\\
&\hspace{-0.5cm}-\!\stackrel{\leftarrow}{\mathcal D}\!\!\!\!\!\phantom{\partial}^\rho(x-\tfrac{z}{2})\mathcal W_{\mathcal C}(x-\tfrac{z}{2},x+\tfrac{z}{2}),
\end{align}
referred to as the mid- and endpoint non-local differential operators, respectively. Using once again Eq. \eqref{Apurefixed}, it is easy to see that the mid- and endpoint non-local pure-gauge covariant derivatives are equivalent $\stackrel{\leftrightarrow}{D}\!\!\!\!\!\phantom{\partial}^\rho_{\text{pure},m}(x-\tfrac{z}{2},x+\tfrac{z}{2})=\stackrel{\leftrightarrow}{D}\!\!\!\!\!\phantom{\partial}^\rho_{\text{pure},e}(x-\tfrac{z}{2},x+\tfrac{z}{2})$. It is therefore possible to unambiguously define a canonical momentum associated with a non-local operator. On the contrary, the mid- and endpoint non-local ordinary covariant derivatives differ
\begin{align*}
&\stackrel{\leftrightarrow}{D}\!\!\!\!\!\phantom{\partial}^\rho_m(x-\tfrac{z}{2},x+\tfrac{z}{2})-\!\stackrel{\leftrightarrow}{D}\!\!\!\!\!\phantom{\partial}^\rho_e(x-\tfrac{z}{2},x+\tfrac{z}{2})=-ig\sum_{\eta=\pm}\nonumber\\
&\quad\int_x^{x+\eta\tfrac{z}{2}} \mathcal W_{\mathcal C}(x-\tfrac{z}{2},s)\,F_{\alpha\beta}(s)\,\mathcal W_{\mathcal C}(s,x+\tfrac{z}{2})\,\frac{\partial s^\alpha}{\partial x_\rho}\,\ud s^\beta
\end{align*}
because of the field-strength tensor, \emph{i.e.} the curvature. No unique kinetic momentum can be associated with a non-local operator. This is particularly obvious when one considers higher $k$-moments. Following the same lines as above, it is easy to show that we can write in general
\begin{align}
&\int\ud^4k\,k^\rho k^\sigma\cdots k^\tau\,\widehat W^{[\gamma^\mu]q}(x,k)=\nonumber\\
&\hspace{1cm} \frac{i}{2}\,\barpsi(x) \gamma^\mu \!\!\stackrel{\leftrightarrow}{D}\!\!\!\!\!\phantom{\partial}^\rho_\text{pure}(x)\!\!\stackrel{\leftrightarrow}{D}\!\!\!\!\!\phantom{\partial}^\sigma_\text{pure}(x)\cdots\!\!\stackrel{\leftrightarrow}{D}\!\!\!\!\!\phantom{\partial}^\tau_\text{pure}(x)\psi(x).
\end{align}
This equation does not suffer from any ambiguity since the pure-gauge covariant derivatives commute, contrarily to the ordinary ones. We therefore conclude that there exists no Wigner operator where the variable $k$ can be interpreted as the (average) kinetic momentum.

\subsection{Orbital angular momentum}

Having at our disposal the Wigner operator, we can simply define the OAM density in phase space as \cite{Lorce:2011kd,Lorce:2011ni}
\begin{equation}\label{OAMformulaunint}
M^{\mu\nu\rho}_{q,\text{OAM}}(x,k)=x^{[\nu}k^{\rho]}\,W^{[\gamma^\mu]q}(x,k).
\end{equation}
Integrating this phase-space density over the momentum simply leads to the OAM density in coordinate space
\begin{equation}\label{OAMformula}
M^{\mu\nu\rho}_{q,\text{OAM}}(x)=\int\ud^4k\,x^{[\nu}k^{\rho]}\,W^{[\gamma^\mu]q}(x,k).
\end{equation}
The (longitudinal) OAM is then defined as usual by the following matrix element
\begin{equation}\label{OAMformulaL}
L_q=\tfrac{1}{2}\,\epsilon^{ij}\,\frac{\langle P,\Lambda|\int\ud^4x\,\delta(x^+)\, M^{+ij}_{q,\text{OAM}}(x)|P,\Lambda\rangle}{\langle P,\Lambda|P,\Lambda\rangle}, 
\end{equation}
where $a^\pm=(a^0\pm a^3)/\sqrt{2}$ and $a^i=a^{1,2}$ are the light-front and transverse components, respectively, and $\epsilon^{ij}$ is the two-dimensional antisymmetric Levi-Civit\`{a} tensor with $\epsilon^{12}=+1$. The proton state with momentum $P$ and light-front helicity $\Lambda$ is normalized as $\langle P,\Lambda|P,\Lambda\rangle=2 P^+(2\pi)^3\,\delta(0^+)\,\delta^{(2)}(\vec 0_\perp)$.

As we have shown in the previous section, the variable $k$ appearing in the Wigner operator can only be interpreted as a canonical momentum. It seems therefore natural to consider that Eq.~\eqref{OAMformulaunint}, and consequently Eqs.~\eqref{OAMformula} and \eqref{OAMformulaL} refer to the \emph{canonical} OAM. In Refs.~\cite{Ji:2012sj,Ji:2012ba}, the authors claim that the \emph{kinetic} OAM can be obtained from Eqs.~\eqref{OAMformula} and \eqref{OAMformulaL} when the Wilson line consists in a direct straight line between the endpoints. We show in the following that, as long as one integrates over the momentum variable $k$, there is actually no contradiction. 

Let us consider the case of the direct straight Wilson line running from the point $x$ to the point $x+z$, and parametrized as usual by the path $s(\lambda)=x+\lambda z$ with $0\leq\lambda\leq 1$. According to Eq.~\eqref{derivativeW}, the derivative of this Wilson line with respect to $z$ gives
\begin{equation}\label{direct}
\frac{\partial}{\partial z^\mu}\mathcal W_{\mathcal C}(x+z,x)=ig\,A_\mu^\text{pure,$x$-FS}(x+z)\,\mathcal W_{\mathcal C}(x+z,x),
\end{equation}
where $A_\mu^\text{pure,$x$-FS}(x+z)$ is the pure-gauge field appearing in the $x$-based Fock-Schwinger GIE defined by the condition $z\cdot A^\text{phys,$x$-FS}(x+z)=0$. Once again, we see that it is only for the component along the path that the ordinary and pure-gauge covariant derivatives do coincide $z\cdot D(x+z)=z\cdot D_\text{pure,$x$-FS}(x+z)$. 

Now, when the momentum variable $k$ is integrated over, one needs only the expressions in the limit $z\to 0$. In that case, we have 
\begin{equation}\label{freefall}
A_\mu(x)=A_\mu^\text{pure,$x$-FS}(x), 
\end{equation}
and consequently $D_\mu(x)=D_\mu^\text{pure,$x$-FS}(x)$. Note that when one changes the point $x$, one also changes the GIE. In other words, we can write in general
\begin{equation}\label{freefall}
A_\mu(x)=\frac{i}{g}\,U(x,y)\frac{\partial}{\partial y^\mu}U^{-1}(x,y)\Big|_{y=x}.
\end{equation}
It is important not to confuse this expression with $A_\mu(x)=\frac{i}{g}\,U(x)\frac{\partial}{\partial x^\mu}U^{-1}(x)$ which would imply that $F_{\mu\nu}(x)=0$. Eq.~\eqref{freefall} just tells us that, at any given point $x$, one can find a gauge transformation such that $\tilde A_\mu(x)=0$ \cite{Cronstrom:1980hj}. This is the analogue of the well-known result in general relativity that one can always find, at any given point $x$, a set of coordinates such that the Christoffel symbols vanish, even if the curvature is nonzero. In some sense, working in the $x$-based Fock-Schwinger gauge corresponds to choosing at the point $x$ the free-fall coordinates in internal space.

In conclusion, even though one cannot interpret the variable $k$ in the Wigner operator as a kinetic momentum, the kinetic OAM can be obtained from the operator \eqref{OAMformula} when the Wilson line consists in a direct straight path. This boils down to that fact that one can always find a gauge transformation such that the gauge field vanishes at a given point.

\section{Light-front gauge-invariant extensions}\label{sec6}

Since the proton spin structure is best probed in high-energy experiments where a parton model is very useful and convenient, it appears that the light-front GIEs, characterized by the contour gauge $A^+=0$, are the most relevant ones. The light-front gauge does not completely fix the gauge freedom, and so in order to define a unique GIE one has to specify also how to fix the residual gauge freedom, typically by imposing boundary conditions. The most popular ones are the advanced ($+$), retarded ($-$) and antisymmetric ($as$) boundary conditions. The canonical OAMs defined in the corresponding light-front GIEs are \emph{a priori} different. It appears however that, because of the time-reversal invariance, they actually coincide. These light-front GIEs have been developed and discussed by Hatta in Refs.~\cite{Hatta:2011ku,Hatta:2011zs}. In the light-front gauge, they simply reduce to the Bashinsky-Jaffe decomposition \cite{Bashinsky:1998if}. The general formalism developed in this paper completes these discussions by treating with greater care the residual gauge freedom and by clarifying some of the arguments.

The advanced and retarded light-front GIEs are typical examples of the contour approach. They are obtained by considering Wilson lines running along the light-front direction to $+\infty^-$ and $-\infty^-$, respectively, where the field is assumed to be a pure gauge. The light-front and transverse components of the different contributions to the gauge field are then given by
\begin{align}
A^+_{\text{con},\pm}(x)&=A^+(x),\\
A^+_{\text{res},\pm}(x)&=A^+_{\text{phys},\pm}(x)=0,\\
A^i_{\text{con},\pm}(x)&=\frac{i}{g}\,\mathcal W_{LF}(x^-,\pm\infty^-)\,\partial^i\mathcal W_{LF}(\pm\infty^-,x^-),\\
A^i_{\text{res},\pm}(x)&=\nonumber\\
&\hspace{-1cm}\mathcal W_{LF}(x^-,\pm\infty^-)\,A^i(\pm\infty^-,\boldsymbol x)\,\mathcal W_{LF}(\pm\infty^-,x^-),\\
A^i_{\text{phys},\pm}(x)&=\nonumber\\
&\hspace{-1.15cm}\int_{\pm\infty^-}^{x^-}\mathcal W_{LF}(x^-,y^-)\,F^{+i}(y^-,\boldsymbol x)\,\mathcal W_{LF}(y^-,x^-)\,\ud y^-,
\end{align}
where we used the notation $\boldsymbol x=(x^+,x^1,x^2)$ and the light-front Wilson line
\begin{equation}
\mathcal W_{LF}(x^-,y^-)\equiv\mathcal P\left[e^{ig\int_{y^-}^{x^-}A^+(y^-,\boldsymbol x)\,\ud y^-}\right].
\end{equation}
On the contrary, the antisymmetric light-front GIE cannot be constructed from a single contour, but can be expressed as an average of the advanced and retarded light-front GIEs
\begin{equation}
A^\mu_{\cdots,as}(x)=\frac{A^\mu_{\cdots,+}(x)+A^\mu_{\cdots,-}(x)}{2},
\end{equation}
where the dots stand for `con', `res' and `phys'.

Since the quark kinetic OAM $L_q=\mathsf L_q-\mathsf L_\text{pot}$ is Stueckelberg invariant, it is sufficient to show that either the quark canonical OAM $\mathsf L_q$ or the potential OAM $\mathsf L_\text{pot}$ is the same in the advanced and retarded light-front GIEs. While Refs.~\cite{Hatta:2011ku,Ji:2012ba,Hatta:2012cs} considered the second option, we prefer the first option because it does not require to go beyond leading twist. As shown in Ref.~\cite{Lorce:2011kd}, the matrix elements of the Wigner operator are related by Fourier transform to the following generalized parton correlator
\begin{align}
&W^{[\Gamma]q}(x,\xi,\vec k_\perp,\vec \Delta_\perp;\eta)=\frac{1}{2}\int\frac{\ud z^-\,\ud^2z_\perp}{(2\pi)^3}\,e^{i(xP^+z^--\vec k_\perp\cdot\vec z_\perp)}\nonumber\\
&\hspace{0.75cm}\langle p',\Lambda'|\overline{\psi}(-\tfrac{z}{2})\Gamma\mathcal W_{LF}(-\tfrac{z}{2},\tfrac{z}{2})\,\psi(\tfrac{z}{2})|p,\Lambda\rangle\Big|_{z^+=0},
\end{align}
where $\Delta=p'-p$ is the momentum transfer, $k^+=xP^+$ and $\Delta^+=-2\xi P^+$ with $P=(p'+p)/2$ the average nucleon momentum, and $\Gamma$ is a Dirac matrix. The Wilson lines simply run along the light-front direction to $\eta\infty^-$ with $\eta=\pm$. This general correlator has been parametrized in Ref.~\cite{Meissner:2009ww} in terms of the so-called generalized transverse-momentum dependent parton distributions (GTMDs). In the vector sector restricted to the leading twist $\Gamma=\gamma^+$, it is parametrized as 
\begin{align}
W^{[\gamma^+]q}&=\frac{1}{2M}\,\overline u(p',\Lambda')\left[F_{1,1}+\frac{i\sigma^{i+}k^i_\perp}{P^+}\,F_{1,2}\right.\nonumber\\
&\left.+\frac{i\sigma^{i+}\Delta^i_\perp}{P^+}\,F_{1,3}+\frac{i\sigma^{ij}k^i_\perp\Delta^j_\perp}{M^2}\,F_{1,4}\right]u(p,\Lambda),
\end{align}
where the GTMDs are complex-valued functions of the variables $(x,\xi,\vec k^2_\perp,\vec k_\perp\cdot\vec\Delta_\perp,\vec\Delta^2_\perp;\eta)$. 

The functions $F_{1,1-3}$ reduce to the usual generalized parton distributions (GPDs) and transverse-momentum dependent parton distributions (TMDs) in the appropriate limits. On the contrary, $F_{1,4}$ appears only at the level of the GTMDs. As shown in Refs.~\cite{Lorce:2011kd,Hatta:2012cs}, it is precisely this leading-twist function which is directly related to the canonical OAM
\begin{equation}\label{OAMGTMD}
\mathsf L_q(\eta)=-\int\ud x\,\ud^2k_\perp\,\frac{\vec k^2_\perp}{M^2}\,F^q_{1,4}(x,0,\vec k^2_\perp,0,0;\eta).
\end{equation}
Now, time-reversal invariance implies that the real part of the GTMDs is $\eta$-even (\emph{i.e.} independent of $\eta$), while the imaginary part is $\eta$-odd. The hermiticity constraint then imposes that the real part of $F^q_{1,4}$ is $\Delta$-even, while the imaginary part is $\Delta$-odd. Since in Eq.~\eqref{OAMGTMD} the GTMD $F^q_{1,4}$ is evaluated at $\Delta=0$, only its real part contributes. This ensures that the canonical OAM is real (as it should be) but also that it does not depend on $\eta$. In other words, the canonical OAM $\mathsf L_q$ defined from the advanced, retarded and antisymmetric light-front GIEs are the same. Similar arguments applied to twist-3 parton correlators show that the potential OAM $\mathsf L_\text{pot}$ is the same in the three light-front GIEs \cite{Hatta:2011ku,Ji:2012ba,Hatta:2012cs}. For an intuitive interpretation, see Ref.~\cite{Burkardt:2012sd}. Note also some recent related discussions \cite{Buffing:2011mj,Buffing:2012sz} treating the path dependence in transverse-momentum dependent correlators.

\section{Conclusion}\label{sec7}

We presented an explicit realization of the Chen \emph{et al.} decomposition of the gauge field into pure-gauge and physical terms. The construction involves Wilson lines and is based on the idea of parallel transport from a reference point. We showed in particular that a change of reference point and/or geodesic induces a change in the explicit separation which can be seen as a Stueckelberg transformation. Paying particular attention to the residual gauge freedom, we proposed a further decomposition of the pure-gauge term into contour and residual terms. 

Then we showed that the momentum variable in the Wigner operator refers to the canonical momentum and not the kinetic momentum. Nevertheless, the kinetic orbital angular momentum can be expressed in terms of the Wigner operator defined with direct straight Wilson lines as long as one integrates over the momentum. Choosing the Wilson lines to run along the light-front direction, our explicit construction simply reduces to the light-front gauge-invariant extensions of the canonical angular momentum discussed in the literature, albeit with a more transparent treatment of the residual gauge freedom. Finally, we showed from twist-2 arguments that the advanced, retarded and antisymmetric light-front canonical orbital angular momenta are the same, confirming the conclusions obtained previously from a twist-3 approach.

\section*{Acknowledgements}
In this study, I greatly benefited from numerous discussions with B. Pasquini during earlier joint work on the Wigner distributions and orbital angular momentum. I am also particularly grateful to A. Metz and L. Szymanowski for many helpful discussions. This work was supported by the P2I (``Physique des deux Infinis'') network.

\end{document}